\documentstyle[12pt]{article}
\include{matex}
\begin{document}
\begin{titlepage}
\pagestyle{empty}
\rightline{CERN-TH.6586/92}
\rightline{FTUV/92-17}
\rightline{IFIC/92-18}
\rightline{July 1992}
\noindent
\begin{center}
{\bf NEW HIGGS SIGNATURES IN SUPERSYMMETRY WITH
SPONTANEOUS BROKEN R PARITY}\\
\vskip 0.4cm
{\bf J. C. Rom\~ao}
\footnote{Bitnet ROMAO@PTIFM}\\
{Centro de F\'{\i}sica da Mat\'eria Condensada, INIC\\
Av. Prof. Gama Pinto, 2 - 1699 Lisboa Codex, PORTUGAL}\\
{\bf F. de Campos}$^1$
\footnote{Bitnet CAMPOSC@EVALUN11 - Decnet 16444::CAMPOSC}\\
and \\
{\bf J. W. F. Valle}$^1$
\footnote{Bitnet VALLE@EVALUN11 - Decnet 16444::VALLE}\\
Theoretical Physics Division, CERN\\
CH-1211 Geneve 23, Switzerland\\
\vskip 0.4cm
{\bf ABSTRACT}\\
\end{center}
\vskip 0.1cm
\noi
Higgs production from $Z$ decays in
supersymmetry with spontaneous broken R parity
proceeds mostly by the Bjorken process as in
the standard model. However, the corresponding production
rates can be weaker than in the standard model (SM),
especially in the low mass region. This will
substantially weaken the Higgs boson mass
limits derived from LEP1.
More strikingly, the main Higgs decay channel is
"invisible", over most of the mass range accessible
to LEP1, leading to events with large missing
energy carried by majorons. This possibility should
be taken into account in the planning
of Higgs boson search strategies not only at
LEP but also at high energy supercolliders.
\vs .1cm
\begin{flushleft}
CERN-TH.6586/92\\
July 1992
\end{flushleft}
\noi
$^1$ {\small Instituto de F\'{\i}sica Corpuscular - C.S.I.C.\\
Dept. de F\'isica Te\`orica, Universitat de Val\`encia\\
46100 Burjassot, Val\`encia, SPAIN}
\end{titlepage}
\setcounter{page}{1}
\pagestyle{plain}
\section{Introduction}

The problem of mass generation is one of the main puzzles
in electroweak physics. Although the Higgs mechanism
\cite{HIGGS} has been suggested more than two decades ago,
it was only very recently with the LEP experiments that one
has started to probe sensitively the nature of the scalar
sector \cite{LEP1}. The existence of supersymmetry (SUSY) at
the electroweak scale is desirable as it can act as a stabilizing
\sym that naturally protects this scale against quantum
corrections associated with superhigh scales.
Although as yet there is no experimental support for
SUSY the above argument has been taken as a strong
motivation to carry out searches at higher energies.
Unfortunately there is no clue as to how SUSY is
realized. The most popular $ansatz$ - called the
minimal supersymmetric standard model (MSSM) \cite{mssm} -
assumes that SUSY is realized in the presence of a discrete
R parity ($R_p$) symmetry under which all standard model
particles are even while their SUSY partners are odd.
This implies that SUSY particles must always be
pair-produced, the lightest of them being stable.
There is great interest in
investigating theories without R parity \cite{fae}.
There are two ways to break $R_p$: explicitly \cite{expl}
and spontaneously \cite{MASI,ZR_RPCHI}. The second
provides a more systematic way to include R parity
violating effects, that moreover automatically respects
low energy {\sl baryon number conservation} and
evades restrictions based on cosmological baryogenesis
arguments \cite{giudice}. Although these could be avoided
even in the explicit $R_p$ breaking scenario \cite{dreiner},
they are certainly avoided in the present model,
inasmuch as the breaking of R parity sets in
spontaneously only as an electroweak scale phenomenon.
Here we concentrate entirely on this second possibility.
There are two cases to consider, depending on
whether lepton number is part of the gauge symmetry
\cite{ZR_RPCHI} or not \cite{MASI}. In the first case
there is a \ZP gauge boson which acquires
mass via the Higgs mechanism at a scale related to
that which characterizes R-parity violation. Here
we focus on the simplest case where $R_p$ is
violated in the absence of an additional gauge
symmetry beyond the minimal \21 structure.
This possibility has been demonstrated
to occur \cite{pot3} in the model suggested in
ref. \cite{MASI} for many suitable values of the low
energy parameters, consistent with observation.
R-parity breaking is driven by
{\sl isosinglet} slepton vacuum expectation
values (VEVS) \cite{MASI}, so that the Goldstone
boson (majoron) associated with spontaneous R parity
breaking is mostly singlet and as a result the $Z$
does not decay by majoron emission, in agreement
with observation \cite{LEP1}. The R-parity breaking
scale typically lies in the phenomenologically
interesting range $\sim 10\:GeV-1\:TeV$,
leading to large rates for the associated
$R_p$ violating effects, which may well
be accessible to experimental test
\cite{ROMA,RPMAJJ,MUTAU}. It also leads to
an interesting way to explain the solar neutrino
data \cite{RPMSW_RPMSWW}.

In this letter we focus on the neutral Higgs sector
of these models and the corresponding implications
for Higgs searches at LEP1. We find that both Higgs
production rates and decay mechanisms may differ
substantially from the SM and MSSM predictions. First,
Higgs production rates can be substantially lower than
in SM and MSSM and, second, its main decay mode can be
invisible,
\beq
h \ra J\;+\;J
\eeq
where $J$ denotes the majoron. This holds
for a wide region of parameters and should also
have important implications for the Higgs search
strategies at high energy hadron supercolliders.

\section{The Model and the Scalar Potential}
\label{ill}

We consider the \21 model defined by the
superpotential
\footnote{All couplings $h_u,h_d,h_e,h_{\nu},h$ are
described by arbitrary matrices in generation space.
Note that we have added some new terms that were not
included in ref \cite{MASI} because they are allowed
by our symmetries. }
terms \cite{MASI}
\bea
\label{P}
h_u u^c Q H_u + h_d d^c Q H_d + h_e e^c \ell H_d +
\hat{\mu} H_u H_d + \\\nonumber
(h_0 H_u H_d - \epsilon^2 ) \Phi +
h_{\nu} \nu^c \ell H_u + h \Phi \nu^c S +
M \nu^c S + M_\Phi \Phi \Phi + \lambda \Phi^3
\eea
The first five terms are the usual ones that define
the $R_p$-conserving MSSM. The fifth term ensures that
electroweak \sym breaking can take place at the tree
level, as in \cite{BFS}. The last five terms involve
\21 superfields $(\Phi ,{\nu^c}_i,S_i)$ carrying a
conserved lepton number assigned as $(0,-1,1)$
respectively. Such singlets arise in several extensions
of the standard model and may lead to interesting
phenomenological signatures of their own
\cite{port,SST1,wein}. Their presence here is
essential in order to drive the spontaneous
violation of R parity and electroweak symmetries in
a phenomenologically consistent way \cite{MASI}.
The bilinear
$H_u H_d$ term plays an
important role in giving more flexibility in the
minimization of the Higgs potential while at the same
time obeying all experimental constraints, especially
the chargino mass limit from LEP. The terms $\Phi^2$,
$\Phi^3$ and $\nu^c S$ do not play any important role
for our present considerations and will be ignored.
We also assume that the coupling
matrices ${h_{\nu}}_{ij}$ and $h_{ij}$ are nonzero
only for the third generation and set
$h_{\nu} \equiv {h_{\nu}}_{33}$ and $h \equiv h_{33}$.
With this assumption we are studying effectively
a one generation model. We are well aware that a
phenomenologically consistent model requires the
presence of flavour nondiagonal couplings \sa
${h_{\nu}}_{23}$, needed in order to ensure that
the massive \nt decays fast enough \cite{fae}
to obey cosmological limits. This has been shown to
be the case due to the existence of the majoron
emission decay channel $\nt \ra \nm + J$. However
for our present purposes the effective one-generation
model approach will be enough.

To complete the specification of the model we give the
form of the full scalar potential along neutral directions
\bea
\label{V}
V_{total}  =
\abs {h \Phi \tilde{S} + h_{\nu} \tilde{\nu} H_u }^2 +
\abs{h_0 \Phi H_u + \hat{\mu} H_u}^2 + \\\nonumber
\abs{h \Phi \tilde{\nu^c}}^2 +
\abs{- h_0 \Phi H_d  - \hat{\mu} H_d +
h_{\nu} \tilde{\nu} \tilde{\nu^c} }^2+
\abs{- h_0 H_u H_d + h \tilde{\nu^c} \tilde{S} - \epsilon^2}^2 +
\abs{h_{\nu} \tilde{\nu^c} H_u}^2\\\nonumber
+ \tilde{m}_0 \left[-A ( - h \Phi \tilde{\nu^c} \tilde{S}
+ h_0 \Phi H_u H_d - h_{\nu} \tilde{\nu} H_u \tilde{\nu^c} )
+ (1-A) \hat{\mu} H_u H_d \right. \\\nonumber
\left. + (2-A) \epsilon^2 \Phi + h.c. \right]
+ \sum_{i} \tilde{m}_i^2 \abs{z_i}^2
+ \alpha ( \abs{H_u}^2 - \abs{H_d}^2 - \abs{\tilde{\nu}}^2)^2
\eea
where $\alpha=\frac{g^2 + {g'}^2}{8}$ and $z_i$ denotes any
neutral scalar field in the theory.

The pattern of spontaneous symmetry breaking of
both electroweak and R parity symmetries has been
studied in \cite{pot3}. There it was demonstrated
explicitly that, for suitable values of the low
energy parameters consistent with observation,
the energy is minimum when both R parity and
electroweak symmetries are spontaneously
broken. Electroweak breaking is driven by the
isodoublet VEVS $v_u = \VEV {H_u}$ and $v_d = \VEV {H_d}$,
assisted by the VEV $v_F$ of the scalar in the singlet
superfield $\Phi$.
The combination $v^2 = v_u^2 + v_d^2$ is fixed by the W mass,
\beq
m_W^2 \simeq \frac{g^2(v_u^2 + v_d^2 )}{2}
\label{mw}
\eeq
while the ratio of isodoublet VEVS determines the important
parameter
\beq
\tan \beta = \frac{v_u}{v_d} \:.
\label{beta}
\eeq
This basically recovers the standard tree level
spontaneous breaking of the electroweak symmetry
in the MSSM. On the other hand the spontaneous breaking
of R parity is driven by nonzero VEVS for the scalar \neus.
The scale characterizing R parity breaking is set by the
isosinglet VEVS
\bea
v_R = \VEV {\tilde{\nu^c}_{\tau}}\\
v_S = \VEV {\tilde{S_{\tau}}}
\eea
where $V = \sqrt{v_R^2 + v_S^2}$ and can lie anywhere in the
range $\sim 10\:GeV-1\:TeV$. A necessary ingredient for the
consistency of this model is the presence of a small
seed of R parity breaking in the $SU(2)$ doublet sector,
\beq
v_L = \VEV {\tilde{\nu}_{L\tau}}\:.
\label{vl}
\eeq
The spontaneous R parity breaking above also entails
the spontaneous violation of $total$ lepton number
(conserved by \eq{P}) leading to the existence of
the majoron, given by,
\beq
\frac{v_L^2}{Vv^2} (v_u H_u - v_d H_d) +
              \frac{v_L}{V} \tilde{\nu_{\tau}} -
              \frac{v_R}{V} \tilde{\nu^c}_{\tau} +
              \frac{v_S}{V} \tilde{S_{\tau}}\:.
\label{maj}
\eeq
Astrophysical considerations \cite{KIM} related to
stellar cooling by majoron emission require a small
value of $v_L \sim 100 \: MeV$, which is also naturally
obtained from the minimization of the Higgs potential
\footnote{
The marked hierarchy in the values of $v_R$ and
$v_L$ follows because $v_L$ is related to a Yukawa
coupling $h_{\nu}$ and vanishes as $h_{\nu} \ra 0$.
This was also shown to have important consequences
for the propagation of solar neutrinos \cite{RPMSW_RPMSWW}.}.

Before entering the discussion of Higgs production
and decay mechanisms we need to complete the
specification of the scalar potential in \eq{V}
by including the relevant radiative corrections.
As has been widely pointed out
\cite{Haber91,Ellis91} there are potentially
important radiative corrections
to the Higgs scalar mass matrices coming from loops
involving heavy quarks and squarks. To take them into
account we followed the effective potential method
used by Ellis, Ridolfi and Zwirner \cite{Ellis91}
for the case of the MSSM.
The one loop effective potential at a scale $Q$
(which we take to be $M_Z$) is given by
\beq
V_{eff}(Q)=V(Q)+ \Delta V(Q)
\eeq
where $V(Q)$ is the tree level potential
given in \eq{V} and
\beq
\Delta V(Q)={ 1 \over 64\pi^2} Str {\cal M}^4 \left( \ln {{\cal M}^2
\over Q^2}- {3 \over 2}\right)
\eeq
In the last equation $Str$ denotes the {\sl Supertrace} and
${\cal M}^2$ is the field-dependent generalized squared mass
matrix for the model. For our purposes it is enough to consider
only the contribution from the bottom and top quarks and their
squarks. In the numerical calculations we also took the same
mass for all the squarks. In this way we calculated $\Delta V(Q)$
as a function of the scalar fields of our model. The radiatively
corrected effective potential obtained in this way was then
minimized following the same procedure used in ref
\cite{pot3}, namely we first solved the extremum
equations
\beq
{\partial V_{eff} \over \partial z_i}=0
\eeq
where $z_i=(H_d,H_u,\tilde{\nu},\tilde{\nu}^c,S,\phi)$.
Then we evaluated the mass squared matrices for the real
and imaginary parts of the fields. A point in parameter
space is a minimum of the potential if all the
eigenvalues of these matrices are positive
(except for the Goldstone modes).

After including radiative corrections we found out that
the basic conclusions of \cite{pot3} remain true in this
case, namely that $R_p$ violation may be energetically
favored over $R_p$ conserving minima. Our results also
indicate that the mass of the lightest CP even state
grows with $m_{top}$ as in the MSSM if this eigenstate
is mostly a doublet, as one would expect. However in our
model this state is often a singlet where the above
considerations do not apply. These points will be further
discussed in sections 3 and 4 where we present our results.

\section{Higgs Production}

The neutral Higgs boson couplings to the $Z$ boson arise in
our model from two sources. In addition to the usual coupling
to $H_u$ and $H_d$ there are couplings to $\tilde{\nu}$. They
come from
\beq
{\cal L}_1=-\frac{i\;g}{2\cos\theta_W}\;Z_\mu\;\left[H_d^{1*}
\stackrel{\leftrightarrow}{\partial^\mu}
H_d^1\;-\;H_u^{2*}\stackrel{\leftrightarrow}{\partial^\mu}\;H_u^2\right]
-\frac{i\;g}
{2\cos\theta_W}\;Z_\mu\;\left[\tilde{\nu}^*\stackrel{\leftrightarrow}
{\partial^\mu}\;
\tilde{\nu}\right]
\label{1}
\eeq
\beq
{\cal L}_2=\frac{g^2}{4\cos^2\theta_W}\;Z_\mu\;Z^\mu\left[H_d^{1*}H_d^1\;+
\;H_u^{2*}\;H_u^2\;+\;\tilde{\nu}^*\;\tilde{\nu}\right]
\eeq
Now we shift the fields according to
\bea
H_d=v_d+\frac{1}{\sqrt{2}}\left[Re(H_d^0)+i\;Im(H_d^0)\right]\\
H_u=v_u+\frac{1}{\sqrt{2}}\left[Re(H_u^0)+i\;Im(H_u^0)\right]\\
\tilde{\nu}=v_L+\frac{1}{\sqrt{2}}\left[Re(\tilde{\nu})+i\;Im(\tilde{\nu})
\right]
\eea
The cubic interactions arising from the ${\cal L}_2$ terms
are rewritten as
\beq
{\cal L}_2=\frac{g}{2\;\cos\theta_W}\;M_Z\;Z_\mu Z_\mu\;\cos\gamma
\left[\cos\beta Re(H_d^0 )+ \sin\beta Re(H_u^0 )+\tan\gamma Re(\tilde{\nu})
\right]
\label{A}
\eeq
where $\tan\gamma=\frac{v_L}{\sqrt{v_u^2 + v_d^2}}$
and $\beta$ is the usual parameter in \eq{beta}.

We now rewrite \eq{A} in the basis of mass
eigenstate scalar bosons. To do this we write the weak  scalar fields  as
a vector $z_i\:\equiv\:(H_d,H_u,\tilde{\nu},\tilde{\nu}^c,\tilde{S},\phi)$.
Then we set
\beq
z_i\:=\:v_i\:+\:\frac{1}{\sqrt{2}}(x_i\:+\:i\:y_i)
\label{zi}
\eeq
where $i=1,...,6$ are the fields defined in the multiplet.
The mass eigenstates are then
 defined by
\beq
H_i = P_{ij}\:x_j \:\:\:\:A_i = Q_{ij}\:y_j
\label{C}
\eeq
where $P$ and $Q$ are orthogonal matrices
\footnote{For simplicity we assume CP conservation.}.

Here $i=1,...,6$ denotes any of the scalar bosons involved.
In the mass eigenstates $H_i$ and $A_i$ the fields
are ordered according to their increasing masses.
In the $0^-$ sector, two of them are massless, one
is the majoron, denoted by $J$, ($A_2$), the other
($A_1$) is the field eaten up by the $Z$.
Substituting \eq{C} into \eq{A} and \eq{1} we find,
respectively,
\beq
{\cal L}_2=\frac{g}{2\;\cos\theta_W}\;M_Z\;Z_\mu Z_\mu\;H_i
\left[\cos\beta P_{i1}+ \sin\beta P_{i2}+\tan\gamma P_{i3}\right]\cos\gamma
\label{ALFA}
\eeq
\beq
{\cal L}_{1}=\frac{g}{2\cos\theta_W} Z_\mu H_i \stackrel{\leftrightarrow}{
\partial^\mu} A_j\;C_{ij}
\label{BETA}
\eeq
where $C_{ij}=P_{i1} \: Q_{j1}- P_{i2}\: Q_{j2}+ P_{i3}\: Q_{j3}$.

In the MSSM limit these expressions reduce to the familiar form where
$P_{11}=-\sin\alpha = - P_{22}$, $P_{12}= \cos\alpha=P_{21}$,
$Q_{21}=\sin\beta$, $Q_{22}=\cos\beta$ and $P_{i3}=0=Q_{i3}$.
As a consequence one recovers in that case the simple result
$C_{12} \ra -\cos\left(\alpha-\beta\right)$,
$C_{22} \ra -\sin\left(\alpha-\beta\right)$ and
$\cos \beta P_{11}+\sin \beta P_{12}= \sin (\beta-\alpha)$.

The interactions in \eq{ALFA} and \eq{BETA} above lead to
the following $Z$ decay channels
\beq
Z \ra H_i \: f \overline{f}
\label{AA}
\eeq
({\sl Bjorken process}) where $f=u,c,d,s,b,e,\mu,\tau$ and
\beq
Z \ra H_i \: A_j
\label{BB}
\eeq
In the MSSM we have $H_1=h$, $H_2=H$ and there is only one physical
pseudoscalar boson $A_2=A$ (there is no majoron in this limit).

It is instructive to compare the predictions on Higgs production
of the present spontaneously broken R parity model with
those of the MSSM. In the latter case there is a complementarity
between the two types of decays in \eq{AA} and \eq{BB}, and this
has been taken as the basis of the experimental analysis
used to place limits on the supersymmetric Higgs
spectrum \cite{Aleph92}.
In the present model there is a breakdown in the sum rule
involving the coupling strengths characterizing the
processes in \eq{AA} and \eq{BB} with respect to
MSSM expectations. As a result overall Higgs production
rates can be weaker than in MSSM. In addition, we have checked
that whenever the "associated" production in \eq{BB}
is kinematically possible, it is dynamically suppressed
because the pseudoscalar Higgs boson A is mostly an
\21 singlet and therefore too weakly coupled to the $Z$.
As a result, Higgs production at LEP proceeds mostly
by the Bjorken process as in the standard model,
except for the possibly suppressed coupling.

Our results for Higgs production via the Bjorken
process in the spontaneously broken R parity model are
summarized in Fig. 1 and 2. For definiteness we have
fixed the value of the R parity breaking scale at
$v_R = v_S = 100 \: GeV$ and $\tan\beta=3$.
The SUSY and $R_p$ parameters have been
randomly varied in the following range:
$-\:250\:<\:\mu\:<\:250\:GeV$,
$30\:<\:M_2\:<250\:GeV$,
$M_1/M_2\:=\:5/3\:\tan^2 \theta_W$,
$m_{\tilde{q}}\:=\:1 TeV$,
$m_{\phi}\:=\:10 \:TeV$,
$10^{-5}\:<\:h_{\nu}\:<\:10^{-1}$.
The remaining parameters have been taken as in
ref. \cite{pot3}.
In all our present analysis we have randomly varied the
relevant unknown parameters over physically
allowed regions only. For example, we have imposed
the lower limit on the lightest chargino mass
$m_{\chi^+}\:>\:45 \:GeV$. Moreover, we have
accepted only those points that are solutions
which minimize the scalar potential of the theory.
In Fig. 1 we show the maximum allowed couplings
characterizing the Bjorken mechanism, versus
the lightest scalar Higgs mass $m_h$
\footnote{The production of the next-to-lightest
scalar Higgs boson $H$ is also possible but less likely.}.
Note that this coupling can be substantially
weaker than in the MSSM because in our model
also $h$ can be predominantly an \21 singlet and
therefore too weakly coupled to the $Z$. This
happens especially for lower mass values, which
are therefore certainly allowed within the
spontaneously broken R parity model
\footnote{
Note that the $Z$ lineshape, \eg the contribution
to $\Gamma_Z^{inv}$ arising from $Z \ra \nu \bar{\nu}h$
is negligible.}.
Indeed, in Fig. 2 we illustrate how the
resulting branching ratio for the
production of leptons in association with
the lightest scalar Higgs $h$ can be much
weaker than expected in the SM.
Limits have been placed on
$BR(Z \ra \ell \bar{\ell} h)$ from the
nonobservation of acoplanar lepton pair events
with the 1990 LEP data sample \cite{Aleph92}
at the level of few $\times 10^{-5}$
(note that the limit becomes weaker for low masses
$m_h \lsim 10\:$ GeV).
Combining all of the experiments one can estimate that
$BR(Z \ra \ell \bar{\ell} h) \lsim 5 \times 10^{-6}$
\cite{dittmar}. From Fig. 2 we see that, even if
the $ZZh$ coupling attains its maximal allowed values,
LEP is barely starting to be sensitive to the
Higgs, as predicted in this model. Similarly
for the branching ratio for the associated
production of quark jets $BR(Z \ra q \bar{q} h)$
as shown in Fig. 2b. However, in the case where
$h$ production is suppressed, the next-to-lightest
of the scalar Higgses, $H$, couples to the $Z$ with
roughly the canonical strength expected for an
isodoublet Higgs. However,
the $H$ production rate will be suppressed with
respect to that in the SM due to the smaller phase
space available. For larger  $h$ masses, the maximum
production rate of the lightest scalar Higgs boson
attains roughly the canonical SM value.

\section{Higgs Decay}

The most novel and characteristic aspect of the
spontaneously broken R parity model is the
existence of the massless pseudoscalar majoron,
which follows from the spontaneous nature of
R parity violation. This implies the presence
of new Higgs decay channels with majoron emission
\footnote{Note that the decays  $A_i \ra JJ$ are
forbidden, since we assume CP conservation},
\bea
H_i \ra J\: J  \\
A_i \ra  H_i \: J
\eea
In order to determine the rates for these decays we need to
evaluate the corresponding trilinear couplings. By using the
definition of $z_i$ we rewrite everything in terms of mass
eigenstates, using \eq{C}. The resulting expressions are
rather long but we have checked that in the MSSM limit
defined by
\beq
h \ra 0 \;\;\;\;\; h_\nu \ra 0 \;\;\;\;\;h_0 \ra 0
\;\;\;\;\;h_0 v_F \ra \mu
\eeq
one gets that the following expression for the
cubic term of the interaction Lagrangian
\beq
{\cal L} =
-\frac{g M_Z}{4\cos\theta_W}
\left[h\:A\:A\: \sin\left(\alpha+\beta\right) \cos2\beta\;
-\;H\:A\:A \:\cos\left(\alpha+\beta\right)\cos2\beta\right]
\label{D}
\eeq
in agreement with MSSM result.

In the present spontaneously broken R parity model the lightest
CP-odd state is the massless majoron. However, because of its
singlet nature, evident from \eq{maj},
all decays involving majoron emission in $Z$ decays
\beq
Z \ra H_i J
\eeq
are automatically suppressed. The situation is quite different
for the majoron emitting Higgs decays. We now proceed to the
determination of the effective $H_i JJ$ coupling relevant
for these decays. Substituting \eq{zi} and \eq{C} in \eq{V}
and keeping only the leading terms in \eq{maj} we find
\bea
\label{E}
g_{H_i JJ}=
-\sqrt{2}\; \left[P_{i6}\left(h^2 v_F + A h m_0 \frac{v_r v_s}{V^2 }
\right)\;+\;P_{i5}\;h^2 v_s \frac{v_R^2 }{V^2 }\;+\;P_{i4} h^2 v_R
\frac{v_S^2}{V^2 } \right. \\\nonumber
+\left. P_{i3}\;h_\nu^2 v_L \frac{v_R^2}{V^2 }\;+\;P_{i2}
\left(-hh_0 v_d \frac{v_R v_S}{V^2 } + h^2_\nu v_u \frac{v_R^2}{V^2 }
\right) \right. \\\nonumber
-\left. P_{i1} hh_0 v_u \frac{v_R v_S}{V^2 } \right]
\eea
The last two terms in \eq{E} indicate that in general
both $h$ and $H$ have unsuppressed coupling to the
majoron.

A simple calculation now enables us to obtain the Higgs
decay rates to $JJ$ and $b\overline{b}$. One finds
\beq
\Gamma\left(H_i \ra b\overline{b}\right)=
\frac{g^2_{H_i b\overline{b}} m_{H_i} }{8\pi}
\eeq
and
\beq
\Gamma\left(H_i\ra JJ \right) =
\frac{ g^2_{H_i JJ}}{32\pi m_{H_i}}
\eeq
so that one finds
\beq
\frac{\Gamma\left(H_i \ra JJ \right)}
     {\Gamma\left(H_i \ra b\overline{b}\right)}=
     \frac{g^2_{H_i JJ}}{m^2_{H_i}}
\frac{1}{g^2_{H_i b\bar{b}}}
\eeq
As a result we expect the "invisible" Higgs decay channel
to be highly dominant, especially for low $m_{H_i}$ values.
We have studied the
attainable values of the invisible Higgs decay branching
ratio $BR_{inv} = BR (h \ra JJ)$ for different $m_h$ values
in the range relevant for LEP studies. We found that the
dominant Higgs decay mechanism is invisible, over most
of the kinematical range available. Since the
majoron is weakly coupled, it will escape undetected.
Thus the $h \ra JJ$ decay will lead to events with
missing energy carried by majorons. For $h$ mass values
in a narrow range close to its maximum ($\sim 80$ GeV
for our choice of $\tan\beta$ and $v_R$) the $h$ decay
pattern approaches that of the SM and $BR_{inv}$ becomes
small. This maximum value of the $h$ mass at which one
changes from the novel regime of dominantly invisible
$h$ decay to the SM regime where $h \ra b \bar{b}$
grows with the value of $\tan \beta$ and $m_{top}$.

\section{Discussion}

We have summarized the results of our study
of the supersymmetric Higgs boson sector of
the model with spontaneous broken R parity
suggested previously \cite{MASI}.
We have shown that in this case Higgs production
from $Z$ decays proceeds mostly by the Bjorken
process as in the standard model. The corresponding
production rates can be weaker than in the standard
model. As a result the LEP1 limit on the lightest
scalar Higgs mass may be substantially weakened,
especially in the low mass region.
More strikingly, the main Higgs decay channel is
likely to be "invisible", over most of the mass
range accessible to LEP1, leading to events with
large missing energy carried by majorons. This
possibility should be considered in the planning
of Higgs boson search strategies not only at
LEP but also at high energy supercolliders.
\vfill
This work was supported by CICYT (Spain) and by a
CNPq fellowship (Brazil). We thank Michael Dittmar,
Rohini Godbole and Roger Phillips for useful discussions.
\newpage
\section*{Figure Captions}
\noindent
{\bf Fig. 1 :}\\
Figures 1a and 1b show the maximum $Z\:Z\:h$ coupling,
normalized with respect to that of the SM, for
$\tan\beta\:=\:3$ and $m_{top}\:=\: 100\:GeV; 150\:GeV$
respectively. All points underneath are possible
in the model, and correspond to the unknown parameters
being randomly varied as specified in the text, including
only physically allowed regions and accepting only
those points which are solutions that minimize
the scalar potential of the theory.
\\
{\bf Fig. 2 :}\\
Figures 2a and 2b show the branching ratios for the
production of the lightest Higgs ($h$) plus leptons
(2a) or quarks (2b) from $Z$ decay. The upper line
is the SM prediction and the lower one is the $maximum$
branching ratio possible in the spontaneously broken
R parity model for the same range of parameters as
in Fig. 1.\\
\newpage
\bibliographystyle{ansrt}
\bibliography{biblio}
\end{document}